\documentstyle[multicol,aps,pre]{revtex}
\begin{document}

\title{Phase-Ordering Dynamics with an Order-Parameter-Dependent 
Mobility: The Large-$n$ Limit}

\author{C. L. Emmott and  A. J. Bray}

\address{Theoretical Physics Group \\
Department of Physics and Astronomy \\
The University of Manchester, M13 9PL, UK}

\maketitle

\begin{abstract}
The effect of an order-parameter dependent mobility (or kinetic 
coefficient), 
given by $\lambda(\vec{\phi})\propto (1-\vec{\phi}^2)^\alpha$, on the
phase-ordering  dynamics of a system described by an $n$-component 
vector order-parameter is addressed at zero temperature in the
large-$n$ limit.  In this limit the system is exactly soluble for both
conserved and non-conserved order parameter; in the non-conserved case the
scaling form for the correlation function and it's Fourier transform,
the structure factor, is established, with the characteristic length
growing as $L\sim t^{1/2(1+\alpha)}$.  In the conserved case, the
structure factor is evaluated and found to exhibit a multi-scaling 
behaviour, 
with two growing length scales differing  by a logarithmic factor: 
$L_1 \sim t^{1/2(2+\alpha)}$ and $L_2 \sim (t/\ln t)^{1/2(2+\alpha)}$.
\end{abstract}

\pacs{64.60.Cn, 82.20.Mj, 05.70.Ln}

\begin{multicols}{2}

\section{Introduction}
 
In this paper we examine the effect of an order-parameter-dependent 
mobility, or kinetic coefficient, on the phase-ordering dynamics of 
a system described by an $n$-component vector order parameter. Both  
conserved and non-conserved order parameters are considered.  For the 
case of a constant (i.e.\ order-parameter independent) mobility/kinetic 
coefficient, both these systems become analytically soluble in the 
large-$n$ 
limit \cite{Coniglio89,Zannetti,Coniglio94}; it is in this limit that we 
now consider the effect of an order-parameter-dependent mobility given by 
$\lambda(\vec{\phi})\propto (1-\vec{\phi}^2)^\alpha$, for models where 
the equilibrium order parameter satisfies $\vec{\phi}^2 = 1$. Thus the 
mobility vanishes in equilibrium, leading to a reduction in the growth 
rate 
of the characteristic length scale, $L(t)$, of the bulk phases. 

The effect of an order-parameter-dependent diffusion coefficient on a
system with a scalar order-parameter has been studied by several authors 
\cite{Lacasta92,Emmott,Puri97} since it has been proposed that for a 
scalar 
order-parameter a mobility of the form $\lambda(\phi)=(1-\phi^2)$ is 
required to accurately model the dynamics of deep quenches \cite{Langer} 
and the effect of an external field \cite{Jasnow}.  Lacasta {\it et. al.} 
\cite{Lacasta92} studied this system numerically using a mobility given 
by 
$\lambda(\phi)=(1-a\phi^2)$.  They found that for $a=1$ the
characteristic length grows as $t^{1/4}$ (in contrast to the
conventional $t^{1/3}$ growth for $a=0$), and for all $a\neq 1$
there is a crossover between $L\sim t^{1/4}$ and $L\sim t^{1/3}$. Similar 
behavior was observed by Puri et al\ \cite{Puri97}. 
This system has been solved exactly in the Lifshitz-Slyosov limit 
\cite{Emmott}
for a more general mobility given by $\lambda(\phi)=(1-\phi^2)^\alpha$; 
in 
this system the system coarsens with growth exponent $1/(3+\alpha)$, 
despite 
the absence of surface diffusion as a coarsening mechanism at late times 
(due to the geometry of the system), and the vanishing of the mobility in 
the 
bulk phases.   

Although a system described by a vector order parameter will have a 
completely 
different morphology form the scalar case (e.g.\ there are no localized
defects for $n>d+1$), it is natural to try to generalise this 
order-parameter 
dependent mobility to the vector case \cite{Corberi}. In this paper 
therefore 
we examine (in sections \ref{chap1:NCOP} and \ref{chap1:COP}) the 
coarsening 
dynamics of an $n$-component vector order-parameter for a general class 
of 
mobilities/kinetic coefficients given by 
$\lambda(\vec{\phi})=(1-\vec{\phi}^2)^\alpha$, where $\alpha\in\Re^+$, 
for 
both the non-conserved and conserved cases.
While these $O(n)$ models are not exactly soluble for general $n$,  
exact solutions can be obtained in the limit $n \to \infty$. 

In section \ref{chap1:NCOP} we consider a non-conserved system with a
vector order parameter.  The scaling hypothesis is established, and the 
exact forms of the two-time correlation function and the structure factor 
are calculated. We find that the characteristic length grows as $L\sim
t^{1/2(1+\alpha)}$.  Due to the absence of defects there is no Porod's 
law: 
the structure factor is Gaussian for all $\alpha$.  

In the conserved case (section \ref{chap1:COP}), the structure factor is 
found to depend on two characteristic lengths, $L_1\sim 
t^{1/2(2+\alpha)}$ 
and $L_2\sim (t/\ln t)^{1/2(2+\alpha)}$, through the form 
$S(k,t) \sim L_1^{d\phi(kL_2)}$. This type of behaviour is termed 
`multiscaling', and the results for $L_1$ and $L_2$ are generalizations 
of 
similar expressions obtained by Coniglio and Zannetti \cite{Coniglio89} 
for the case of a constant mobility.  Indeed, as expected, all the 
results 
of this paper reduce to the established constant $\lambda$ results when 
$\alpha$ is set to zero.
 
We conclude with a summary and discussion of the results.

\section{The Non-Conserved O(n) Model}
\label{chap1:NCOP}

 The dynamics of a non-conserved  vector order parameter are described by 
the
phenomenological time-dependent Ginzburg-Landau equation \cite{Review},
\begin{equation}
{\partial\phi_i\over\partial t} = 
- \lambda (\vec{\phi}^2 ){\delta F[\vec{\phi}]\over \delta \phi_i}
=\lambda (\vec{\phi}^2 )
\left( {\mbox{\boldmath $\nabla$}}^2\phi_i -
{\partial V(\vec{\phi}^2)\over\partial \phi_i} \right),
\label{on:ncop}
\end{equation}
where $ V(\vec{\phi}^2) $ is the potential energy term in the
Ginzburg-Landau free energy functional, and is invariant
under global rotations of $\vec{\phi}$.
  In the following calculation, the conventional choice is made for the
form of the
potential:
\begin{equation}
 V(\vec{\phi}^2) = {(1-\vec{\phi}^2)^2 \over 4 },
\end{equation}
and the order-parameter-dependent kinetic coefficient is given by 
$\lambda(\vec{\phi})=(1-\vec{\phi}^2)^\alpha$.

 In the limit $ n\to \infty$ equation (\ref{on:ncop}) may be simplified by
making the following substitution, 
\begin{equation}
\vec{\phi}^2 = \lim_{n\to\infty}
  \biggl( \sum_{j=1}^n\phi_j^2 \biggr) =\,n\, <\!\phi_k^2\!> 
= \langle \vec{\phi}^2 \rangle\ .
\label{on:simply3}
\end{equation}
\noindent where $<\!...\!>$ represents an ensemble average. 
Defining $a(t)$ by the equation $a(t) = (1 -
n<\!\vec{\phi}^2\!> )$, equation (\ref{on:ncop}) then reduces to
\begin{equation}
{\partial\phi_i\over\partial t} = a^{\alpha}(t)
\biggl( {\mbox{\boldmath $\nabla$}}^2 +
a(t) \biggr)\phi_i.
\end{equation}
If we now take the Fourier transform, this equation can easily be
solved to give
\begin{equation}
\phi^{(i)}_{\bf k}(t) = \phi^{(i)}_
{\bf k}(0)  
\exp (-k^2b(t) + c(t)),
\label{on:ncop3}
\end{equation}
where $b(t) = \int_0^tdt'\,a^{\alpha}(t')$ and 
$c(t) = \int_0^tdt'\,a^{1+\alpha}(t')$.  
On substituting equation (\ref{on:ncop3})  back into the definition
 of $a(t)$ we find
\begin{equation}
a(t)=1-\Delta\exp[2c(t)]\sum_{\bf k}\exp[-2 k^2b(t)],
\label{on:ncop13}
\end{equation}
where we have used the conventional choice for the initial conditions,
\begin{equation}
 <\!\phi^{(i)}_{\bf k}
  \phi^{(j)}_{-{\bf k}'}\!> =
 \left({\Delta\over n}\right)\,
\delta_{ij}
  \,\delta_{{\bf k}\,
{\bf k}'}.
\label{on:init}
\end{equation}
Using the fact that $ \sum_{\bf k} \exp(-2k^2b(t))=
(8\pi b(t))^{-d/2}$ in equation (\ref{on:ncop13}) we obtain
\begin{equation}
a(t)=1-\Delta [8\pi b(t)]^{-d/2}  \exp[2c(t)].
\label{abc}
\end{equation}
Since we are mainly interested in late times, we now solve this
equation self-consistently to obtain the large-$t$
result for $b(t)$ and $c(t)$.  In order to make progress we
make the assumption that at late times $a(t)\ll 1$, and hence the term
on the left-hand side
of equation (\ref{abc}) may be neglected.  The validity of this 
assumption 
will be proved {\it a posteriori}. Thus we wish to solve  
\begin{equation}
\Delta [8\pi b(t)]^{-{d\over 2}} \exp [2c(t)] = 1.
\label{on:aconsistant3}
\end{equation}  
Differentiating this expression with respect to time gives the
following relation,
\begin{equation}
\dot{c}(t)={d\,\,\dot{b}(t)\over 4\,b(t)}.
\label{nc:above}
\end{equation}
Substituting the derivatives of $b(t)$ and $c(t)$, which are
given by: 
\begin{eqnarray}
\dot{b}(t)&=&a^{\alpha}(t),\label{on:dotb} \\
\dot{c}(t)&=&a^{1+\alpha}(t),\label{on:dotc}
\end{eqnarray}
into equation (\ref{nc:above}),  we find that
\begin{equation}
 b(t) = {d\over 4 a(t)}.
\label{on:ba}
\end{equation}
If we now differentiate again, we obtain a simple differential equation 
for
$a(t)$, and from this we find that the large-$t$ behaviour of $a(t)$
is given by
\begin{equation}
 a(t) \sim  \left( {4(1+\alpha) t \over d} \right)^{-{1\over 1+\alpha}}.
\end{equation}
Hence it can clearly be seen that the assumption that $a(t) \ll 1$ at late
times is justified. 

Using this result together with equations (\ref{on:aconsistant3})
and (\ref{on:ba}), we find that
\begin{eqnarray}
b(t)&\sim & \sigma t^{1/(1+\alpha)}, \label{ncop:b}\\
c(t)&\sim &{d\over 4(1+\alpha)}\ln\left({t\over t_0}\right),
\label{ncop:c}
\end{eqnarray}
where
\begin{eqnarray}
\sigma &=& (1+\alpha)^{1\over 1+\alpha}\,
\biggl({d\over 4} \biggr)^{\alpha\over 1+\alpha},\\
t_0 &=& {1\over\alpha+1} \biggl({4\over d} \biggr)^{\alpha}\,
 \biggl({\Delta^{2\over d} \over 8\pi}\biggr)^{1+\alpha}.
\end{eqnarray}

We are now in a position to evaluate the expression for the Fourier
  transform of the order parameter at large $t$.  Substituting equations
 (\ref{ncop:b}) and (\ref{ncop:c}) into equation  (\ref{on:ncop3}) we 
find that
\begin{equation}
\phi^{(i)}_{\bf k}(t) = 
\phi^{(i)}_{\bf k}(0)  
\biggl({t\over t_0}\biggr)^{d\over 4(1+\alpha)}
\exp\bigl(-\sigma k^2t^{1\over 1+\alpha} \bigr).
\label{nc:phi}
\end{equation}
Using this result, we can evaluate the two-time structure factor
and the correlation function.  These are given by:
\begin{eqnarray}
S({\bf k},t_1,t_2) &=& (8\pi \sigma)^{d/ 2}(t_1t_2)^{d/ 
4(1+\alpha)}\nonumber\\
&&\times
\exp\left(-\sigma k^2( t_1^{1/(1+\alpha)}+  t_2^{1/(1+\alpha)} )\right),\\
C({\bf r},t_1,t_2) &=& \left({4(t_1t_2)^{1/(1+\alpha)}\over
(t_1^{1/(1+\alpha)}+  t_2^{1/(1+\alpha)})^2 }   \right)^{d/4}\nonumber\\
&&\times
\exp\left({-x^2\over 4\sigma 
(t_1^{1/(1+\alpha)}+  t_2^{1/(1+\alpha)} )}\right),
\end{eqnarray}
which, in the equal time case, reduce to the following expressions:
\begin{eqnarray}
S({\bf k},t) &=& (8\pi \sigma)^{d/ 2}t^{d/ 2(1+\alpha)}
\exp\left(-2\sigma k^2t^{1/(1+\alpha)} \right),\\
C({\bf r},t) &=& 
\exp\left(-{x^2\over 8 \sigma t^{1/(1+\alpha)} }\right).
\end{eqnarray}
These results exhibit the expected scaling forms,
with the characteristic length scale growing as $L \sim 
t^{1/2(1+\alpha)}$.
The structure factor has a Gaussian form, without the power-law tail 
predicted by Porod's law.  This is a direct consequence of the absence 
of defects in the system.

If we now look at the two-time correlation function in the limit
$t_1 \gg t_2$, we find that
\begin{equation}
C({\bf r},t_1,t_2) =
 \left[4\left({t_2\over t_1}\right)^{1/(1+\alpha)}  \right]^{d/4}
\exp\left(-{x^2\over 4\sigma t_1^{1/(1+\alpha)}}\right).
\end{equation}
Comparing this with the scaling form \cite{Review}
$C({\bf r},t_1,t_2)=(L_2/L_1)^{\overline{\lambda}}\,h(r/L_1)$, we obtain
the result, $\overline{\lambda} =d/2$, independent of $\alpha$. 

It is also interesting to compare the response function, 
$G({\bf k},t) = \langle d\phi^{(i)}_{\bf k}(t)/d\phi^{(i)}_{\bf 
k}(0)\rangle$, 
with the structure factor $S({\bf k},t,0)$, i.e.\ with the correlation of 
$\phi^{(i)}_{\bf k}(t)$ with its $t=0$ value. Using equation 
(\ref{nc:phi}) we find that:
\begin{eqnarray}
S({\bf k},t,0)&=& 
\Delta
\left({t\over t_0}\right)^{d\over 4(1+\alpha)}
\exp\left(-\sigma k^2t^{1\over 1+\alpha} \right),\\
G({\bf k},t)&=& \left({t\over t_0}\right)^{d\over 4(1+\alpha)}
\exp\left(-\sigma k^2t^{1\over 1+\alpha} \right),
\end{eqnarray}
which verifies the relation
$S({\bf k},t,0)=\Delta G({\bf k},t)$.  Note that this is an exact result 
valid
beyond the large-$n$ limit; this may be proved by integration by parts
on the Gaussian distribution for $\{\phi_{\bf k}(0)\}$ \cite{Kissner92} .

\section{The Conserved O(n) Model}
\label{chap1:COP}

 The dynamics of a system described by a 
conserved vector order parameter are modelled by the
Cahn-Hilliard equation \cite{Review},
\begin{eqnarray}
{\partial\phi_i\over\partial t} &=& \mbox{\boldmath $\nabla$} . 
\left(\lambda
(\vec{\phi}^2){\mbox{\boldmath $\nabla$}}
 \left({\delta F[\vec{\phi}]\over\delta\phi_i}\right)\right)\nonumber 
\\&=&
    \mbox{\boldmath $\nabla$} . \left(\lambda
(\vec{\phi}^2){\mbox{\boldmath $\nabla$}}
 \left( -{\mbox{\boldmath $\nabla$}}^2\phi_i +
{\partial V(\vec{\phi}^2)\over \partial\phi_i}\right)\right),
\label{on:cop}
\end{eqnarray}
where we make the same choice for the potential as before,
$ V(\vec{\phi}^2) = {1\over 4}(1-\vec{\phi}^2)^2$.  Following the 
method of the previous calculation, $\vec{\phi}^2 $ is eliminated using
equation (\ref{on:simply3}), therefore equation (\ref{on:cop}) reduces
to
\begin{equation}
{\partial\phi_i\over\partial t} = -a^{\alpha}(t)
\biggl({\mbox{\boldmath $\nabla$}}^4\phi_i
+ a(t) {\mbox{\boldmath $\nabla$}}^2\phi_i \biggr),
\end{equation}
where $a(t)$ is defined as before.
  Taking the Fourier transform and solving the resulting differential
equation yields
\begin{equation}
\phi^{(i)}_{\bf k}(t) = \phi^{(i)}_{\bf k}(0)
\exp\left(- k^4b(t) + k^2c(t)
\right),     
\label{on:phi}
\end{equation}
 where $b(t) $ and $c(t) $ are defined as for the
non-conserved case.  Substituting this back into the formula for $a(t)$ 
and
using the random initial conditions given by equation (\ref{on:init}) 
gives
\begin{equation}
a(t) = 1 - \Delta\,
\sum_{\bf k}\exp\left(- 2k^4b(t) +2 k^2c(t)\right).
\label{on:shit}
\end{equation} 

 To make further progress we again assume that at large $t$, 
$a(t) \ll 1$.  This is checked for self-consistency later in
the calculation.  The sum over ${\bf k}$ is converted to an 
integral and,
using the change of variables 
\begin{equation}
{\bf x} = \biggl({b(t)\over c(t)}
	\biggr)^{1\over 2}\,{\bf k},
\end{equation}
 equation (\ref{on:shit}) becomes
\begin{eqnarray}
1&=& {\Delta\over 2^{d-1}\pi^{d/2}\Gamma (d/2)}
\left({\beta(t)\over  b(t)}\right)^{d\over 4}\,\nonumber\\ &&\times
\int_0^{\infty} dx\,x^{d-1}\exp\left( 2\beta(t)(x^2-x^4)\right), 
\label{on:steep} 
\end{eqnarray}  
where 
\begin{equation}
\beta(t)=c^2(t)/b(t).
\label{on:defbeta}
\end{equation}

We now make an additional assumption (also to be verified 
{\it a posteriori}) that $\beta(t)\to\infty $ as $t\to\infty$; 
the integral on the left-hand side of equation (\ref{on:steep}) can then 
be 
evaluated by the method of steepest descents.  Therefore, equation 
(\ref{on:steep}) finally simplifies to
\begin{equation}
{\Delta\,\beta(t)^{-1/2}\over 2^{3d/2}\pi^{(d-1)/2}\Gamma(d/2)}  
\left({\beta(t)
\over b(t)}\right)^{d/4}
\,\exp[\beta(t)/2] = 1.
\label{on:log}
\end{equation}

We now  solve this equation asymptotically, obtaining expressions for
$a(t)$, $b(t)$ and $\beta(t)$ at late times.  On
 taking the  logarithm of equation (\ref{on:log}) we find that
\begin{equation}
\beta(t) \simeq {d\over 2}\ln b(t) 
+ \left({2-d \over 2}\right)\ln[\ln b(t)].
\label{on:beta1}
\end{equation}
Using the definition of $\beta(t)$ (equation
(\ref{on:defbeta})) in equation (\ref{on:beta1}) we obtain an
equation for $c(t)$, which when differentiated, gives (to
leading order)
\begin{equation}
\dot{c} (t)\simeq 
\left({d\ln b(t)\over 8b(t) } \right)^{1/2}\dot{b} (t).
\end{equation}
If we now substitute for the derivatives of $b(t)$ and 
$c(t)$ from equations (\ref{on:dotb}) and (\ref{on:dotc}) respectively, 
we find
that
\begin{equation}
a^{\alpha}(t)=\dot{b} (t)=
\left({d\ln b(t)\over 8b(t) } \right)^{\alpha/2},
\end{equation}
which has the asymptotic solution
\begin{equation}
 b(t) \simeq 
\left({(2+\alpha)t\over 2}\right)^{2/(2+\alpha)}
\left({d\ln t \over 4(2+\alpha)}\right)^{\alpha /(2+\alpha)}.
\label{on:asymb}
\end{equation}

If we now differentiate this expression once more, we obtain the
asymptotic behaviour of $a(t)$,
\begin{equation}
a(t)\simeq 
\left({d \ln t\over 2(2+\alpha)^2t}\right)^{1 /(2+\alpha)}
\left(1 + {1\over 2\ln t} \right),
\end{equation}
and clearly $a(t) \ll 1$ at late times, justifying one of
our initial assumptions.

On substituting equation (\ref{on:asymb}) into equation
 (\ref{on:beta1}), we obtain 
\begin{equation}
\beta(t) \simeq {d \over 2+\alpha}\ln t
+\left({2+\alpha-d\over 2+\alpha }\right)\ln(\ln t).
\label{on:asymbeta}
\end{equation}
 We see that as $t\to\infty$, $\beta(t)\to\infty $, justifying
the application of the method of steepest descents to the
integral in equation (\ref{on:steep}). Thus both our initial
assumptions are satisfied.

We are now in a position to evaluate the expression for 
$\phi^{(i)}_{\mbox{\scriptsize{\bf k}}}(t)$.  
 Completing the square in
the exponent on the right-hand side of equation (\ref{on:phi}) gives,
\begin{eqnarray}
\phi^{(i)}_{\bf k}(t)&=&\phi^{(i)}_{\bf k}(0)
\exp\Biggl[{\beta(t)\over 4}
\nonumber\\ &&\qquad\qquad-  {\beta(t)\over 4}  \left(1-
2\left({b(t)\over\beta (t)}\right)^{1/2}k^2 \right)^2
\Biggr].
\end{eqnarray}
Substituting for  $b(t) $ and
 $\beta(t)$, from equations (\ref{on:asymb}) and
(\ref{on:asymbeta}) respectively, gives
\begin{equation}
\phi^{(i)}_{\bf k}(t)\simeq \phi^{(i)}_{\bf k}(0)
\left(\ln t\right)^{2+\alpha -d\over 4(2+\alpha)}
t^{{d\over 4(2+\alpha)}\phi (k/k_m)},
\end{equation}
where
\begin{equation}
k_m=\left({d\ln t\over 2(2+\alpha)^2 t}
\right)^{1\over 2(2+\alpha)}
\end{equation}
is the position of the maximum in the structure factor,  
and $\phi(x)=1-(1-x^2)^2$.

The structure factor is therefore given by
\begin{equation}
S({\bf k},t)\simeq\Delta
\left(\ln t\right)^{2+\alpha -d\over 2(2+\alpha)}
t^{{d\over 2(2+\alpha)}\phi (k/k_m)}.
\end{equation}
>From this expression it is self-evident that the structure factor does 
not have the
conventional scaling form $S({\bf k},t)\sim L^dg(kL)$.  In 
 this system there are  two different length scales, $L_1$ and $L_2$, 
which 
which differ only by a logarithmic factor and are given by
\begin{eqnarray}
L_1 &\sim & t^{1/2(2+\alpha)}, \\
L_2 &\sim & k_m^{-1} = \left({t\over \ln t} \right)^{1/2(2+\alpha)}.
\end{eqnarray}

The structure
factor is therefore of the form $S({\bf k},t)\sim L_1^{d\phi(kL_2)}$
with an additional logarithmic correction factor,
$\left(\ln t\right)^{2+\alpha -d\over 2(2+\alpha)}$;
the exponent depends continuously on a scaling variable.
This type of behaviour is called `multiscaling', and was first noted by 
Coniglio and Zannetti for the case $\alpha=0$ \cite{Coniglio89}. Note 
that 
the $\alpha$-dependence enters through the length scales $L_1$ and $L_2$, 
while the function $\phi(x)$ is independent of $\alpha$.

\section{Discussion and Conclusions}
In this paper we have considered the effect of an 
order-parameter-dependent mobility/kinetic coefficient, given by
$\lambda(\vec{\phi})=(1-\vec{\phi}^2)^\alpha$, on a system described by
an $n$-component vector order parameter. Exact results have been obtained 
in the large-$n$ limit, a limit which despite its limited applicability 
to physical systems has been widely studied as one of the few exactly 
soluble models of phase-ordering kinetics
\cite{Coniglio89,Zannetti,Coniglio94,Newman90b,Newman90a,Humayun92a,Humayun92b,Kissner93}.
All the results obtained  reduce to the
expected constant $\lambda$ results when $\alpha$ is set to zero.  

In the non-conserved system, the correlation function and its Fourier
transform, the structure factor, were explicitly calculated and found
to be of the expected scaling form, with the characteristic length 
growing as $L\sim t^{1/2(1+\alpha)}$.  The 
order-parameter-dependent kinetic coefficient slows down the rate of 
domain 
coarsening; the result reduces to the familiar $t^{1/2}$ growth for the 
case 
$\alpha =0$ \cite{Review,Newman90a}. 
The result $\overline{\lambda}=d/2$, independent of $\alpha$, was 
established 
from the two-time correlation function $C({\bf r},t_1,t_2)$ in the regime
$t_1\gg t_2$, and the relation $S({\bf k},t,0)=\Delta G({\bf k},t)$, 
relating 
the correlation with, and the response to, the initial condition was
verified.  The equal-time correlation functions and structure factor are 
Gaussian.
  
The system with a conserved order parameter was found to exhibit
a more unusual behaviour.  In this system, the structure factor does 
not have the conventional scaling form and is dependent on {\it two} 
scaling
lengths, $t^{1/2(2+\alpha)}$ and $k_m^{-1} \sim (t/\ln 
t)^{1/2(2+\alpha)}$, 
where $k_m$ is the position of the maximum in the structure factor. 
This type of behaviour was first discovered in a phase-ordering system by 
Coniglio and Zannetti \cite{Coniglio89}, for the $\alpha =0$ case. 
For $\alpha = 0$ this behaviour is a consequence of the non-commutativity
of the large-$n$ and large-$t$ limits, as demonstrated within a soluble 
approximate model by Bray and Humayun \cite{Humayun92a}. They 
demonstrated 
that for finite $n$, in the limit $t\to\infty$, conventional scaling is 
found whereas if the $n\to\infty$ limit is taken first (at finite $t$), 
the Coniglio and Zannetti result \cite{Coniglio89} is recovered.
At large, but finite $n$, multiscaling behaviour is found at 
intermediate times, with a crossover to simple scaling behaviour 
occurring at late times \cite{Humayun92a,CZ,CCZ}. We anticipate that a 
similar 
crossover to simple scaling at late times will occur for any $\alpha$ for 
large but finite $n$, leaving a single growing length scale 
$L \sim t^{1/2(2+\alpha)}$, but an explicit demonstration of this goes 
beyond the scope of the present work. 

Note that all the results presented above have been derived 
in the absence of thermal noise, so these results are strictly valid only 
for quenches to $T=0$.  However, since we do not
expect temperature to be a relevant variable \cite{Review,Newman90b},
qualitatively similar results should be obtained for quenches to $T>0$ 
(but 
$T<T_c$), at least for nonconserved dynamics (with $n$ finite or 
infinite) or 
conserved dynamics with finite $n$ \cite{CCZ}.

\section{Acknowledgements}
This work was  supported by EPSRC (United Kingdom).

\end{multicols}

\end{document}